\documentclass{article}
\usepackage{geometry}
\usepackage{amsmath}
\usepackage{caption}
\usepackage{subcaption}
\usepackage{natbib}
\RequirePackage{algorithm2e}
\RequirePackage{amsthm,amsmath,amsfonts,amssymb}
\RequirePackage{amssymb}
\RequirePackage{graphicx}
\RequirePackage{hyperref}
\RequirePackage{pifont}
\usepackage{booktabs}
\usepackage{authblk}
\usepackage{fancyhdr}
\newcommand{\cmark}{\ding{51}}%
\newcommand{\xmark}{\ding{55}}%
 \usepackage{graphicx}
\title{PDE-Based Bayesian Hierarchical Modeling for Event Spread,\\
with Application to COVID-19 Infection}

\author[1]{Mengqi Cen}
\author[2]{Xuejing Meng}
\author[3]{X. Joan Hu}
\author[4]{Juxin Liu}
\author[5]{Jianhong Wu}

\affil[1]{\raggedright Department of Population Medicine, Harvard Pilgrim Health Care Institute, Boston, Massachusetts, USA\\
\texttt{mengqi\_cen@hphci.harvard.edu}}
\affil[2]{\raggedright School of Statistics and Mathematics and Collaborative Innovation Center of China Pilot Reform Exploration and Assessment Hubei Sub-Center, Hubei University of Economics, Wuhan, China.\\
\texttt{mengxuejing18@163.com}}
\affil[3]{\raggedright Corresponding author, Department of Statistics and Actuarial Science, Simon Fraser University, Burnaby, Canada.\\
\texttt{joanh@stat.sfu.ca}}
\affil[4]{\raggedright Department of Mathematics and Statistics, University of Saskatchewan, Saskatoon, Canada.\\
\texttt{jul086@mail.usask.ca}}
\affil[5]{\raggedright Department of Mathematics and Statistics, York University, Toronto, Canada.\\
\texttt{wujh@mathstat.yorku.ca}}

\date{}

\begin{document}
\maketitle
\begin{abstract}

{We extended the Wikle's Bayesian hierarchical model based on a diffusion-reaction equation~\citep{ wikle2003hierarchical} to investigate the COVID-19 spatio-temporal spread events across the USA from Mar 2020 to Feb 2022. Our model incorporated an advection term to account for the intra-state spread trend. We applied a Markov chain Monte Carlo (MCMC) method to obtain samples from the posterior distribution of the parameters. We implemented the approach via the collection of the COVID-19 infections across the states overtime from the New York Times. Our analysis shows that our approach can be robust to model misspecification to a certain extent and outperforms a few other approaches in the simulation settings. Our analysis results confirm that the diffusion rate is heterogeneous across the USA, and both the growth rate and the advection velocity are time-varying.}
\end{abstract}

\section{Introduction}

Coronavirus disease (COVID-19) emerged in December 2019, and has rapidly spread into a pandemic of the world since then. Millions of confirmed cases have been reported globally. The United States, for example, has more than 80 million infected people as of Mar 31, 2022~\citep{data}.

In response to the ongoing pandemic, numerous studies have been conducted to understand and predict the spread of COVID-19. The most commonly used models are the compartmental models. The population is assigned to compartments according to the infectious status, and the transfer rates between compartments are estimated. For instance, the susceptible-infectious-recovered (SIR) model with time-varying transmission rates is used to predict the epidemic trend of COVID-19~\citep{wangping2020extended}. Cooper et al. predict the spread of the virus using the extended SIR model with time-varying total and susceptible population~\citep{cooper2020sir}. Other models, such as the Susceptible-Infected-Recovered-Deceased (SIRD), Susceptible-Exposed-Infectious-Removed (SEIR) and Susceptible-Exposed-Infectious-Quarantined-Removed (SEIQR) models, are developed by including more compartments~\citep{bousquet2022deep,delli2022hybrid,sun2020transmission}. These models are based on ordinary differential equations (ODEs), and are only applicable to studying disease processes over time. Given that most regions do not follow complete isolation, modeling the spread in both space and time is desirable and can be more realistic.

To investigate event spatio-temporal spread, the diffusion process is widely used~\citep{ wikle2003hierarchical,brauer2019mathematical}. Consider the general advection-diffusion partial differential equation (PDE):
\setlength{\abovedisplayskip}{3pt}
\begin{equation}\label{wiklepde}
    \frac{\partial}{\partial t}u(t;s)=\nabla \cdot \delta(s) \nabla u(t;s) +\zeta u(t;s),
\end{equation}

\noindent where $u(t;s)$ is the spatio-temporal process of interest with $s=(x,y)^T \in \mathcal{S}$ at time $t \in \mathcal{T}$, $\delta(s)$ is the spatially varying diffusion coefficient and $\zeta$ is a growth coefficient. Here $\bigtriangledown=(\frac{\partial}{\partial x},\frac{\partial}{\partial y})^{T}$ is the gradient operator, and for a vector field $F=(F^x,F^y), \bigtriangledown \cdot F=\frac{\partial F^x}{\partial x}+\frac{\partial F^y}{\partial y}$ is the divergence operator. This PDE has been applied to model the spread of COVID-19: $\delta$ represents the speed of spread between neighbouring regions, and $\zeta$ indicates the speed of spread within a region~\citep{wang2020using,yamamoto2021quantifying}. 

Motivated by patterns shown in the American COVID-19 data from \emph{the New York Times}~\citep{data}, we consider an advection-diffusion-reaction PDE in this paper. The extended PDE is widely used in many areas especially in environment and ecology~\citep{hundsdorfer2007numerical,sigrist2015stochastic}.

When analytic solutions are absent, the finite difference method (FDM) is frequently used to approximate the solution of a PDE~\citep{smith1985numerical}. It solves differential equations by replacing the derivatives with their finite difference approximations. First-order forward differences in time and centred differences in space are extensively used to approximate spatio-temporal processes; see, for example~\cite{lu2020nonlinear} and~\cite{williams2019rise}. This paper uses FDM to approximate the solution of the advection-diffusion-reaction PDE, which defines the latent spatio-temporal process underlying the event spread.

This paper is structured as follows. Section \ref{chap:1.2} starts with a descriptive analysis of the American COVID-19 data. Section \ref{chaphba} elaborates the PDE-based Bayesian hierarchical model and presents the computational techniques using MCMC methods for parameter estimation. In Section \ref{chptnume}, the proposed approach is assessed by simulation studies. It is followed by COVID-19 data analysis. Section \ref{chptdiscussion} includes conclusions and comments regarding future investigations.

\section{American COVID-19 Data from Jan 21, 2020 to Mar 31, 2022}\label{chap:1.2}

We downloaded the records of American COVID-19 infections from \emph{the New York Times} on April 1, 2022. The data were gathered from the local health agencies’ reports, including the number of daily cumulative lab-confirmed cases in each American county from Jan 21, 2020, to Mar 31, 2022.

Figure \ref{1.1} shows the confirmed case numbers of the mainland USA during the selected periods. Each point is located at the centre of a county, and the size of the point is proportional to the county's case counts. The plot suggests that the spread does not have the same diffusion rate in all directions. Thus, the PDE in (\ref{wiklepde}) is not sufficient to explain the spread of the virus. An advection term should be included to explain the tendency of the virus for movement in a specific direction.

Figure 1 in the Supplementary Materials presents the daily cases with a 7-day moving average in the USA. There were three large spikes in the confirmed case numbers over the past two years. The trend of case counts is clearly varying over time. It suggests that a more generalized model with time-varying parameters should be considered.

\section{Hierarchical Bayesian Analysis}\label{chaphba}
\subsection{PDE-Based Bayesian Hierarchical Modeling}\label{chap:2.1}

We introduce a hierarchical Bayesian model to explore the event spatio-temporal spread on a regular grid. The hierarchical Bayesian modeling consists of three stages: the first stage specifies the distribution of the data, the second stage formulates a latent process, and the last stage assigns the prior distribution to the parameters in the model.

\subsubsection{Modeling Event Counts}

Let $n(t;s)$ be the total number of events of interest in location $s=(x,y)^T \in \mathcal{S}$ at time $t \in \mathcal{T}$. Assume $n(t;s)$ follows a Poisson distribution with mean $\lambda(t;s):$

\begin{equation}\label{eq1}
    n(t;s)\mid \lambda(t;s) \sim Poisson (\lambda(t;s)).
\end{equation}

Here we assume that $n(t;s)$ and $n(t';s')$ are independent conditional on their means $\lambda(t;s)$ and $\lambda (t';s')$ for either $t \neq t'$ or $s \neq s'$.

\subsubsection{Latent Process Specification}
Since the Poisson intensity can take only positive real values, we consider $u(t;s)=log(\lambda(t;s))$. To take other important factors into account, one may consider $u(t;s)=log(\lambda(t;s))-log(c(t;s))$, where $c(t;s)$ is the population size at location $s$ at time $t$ in COVID-19 application. We specify $u(t;s)$, a latent spatio-temporal process, in the following.

To describe the latent spatio-temporal process underlying a collection of bird migration data, Wikle considers a diffusion-reaction equation with a space-varying diffusion rate as shown in (\ref{wiklepde}). This partial differential equation (PDE) is widely used in modeling ecological and epidemic processes, for instance, the expansion of Eurasian Collared-Dove~\citep{hooten2008hierarchical} and the indoor spread of coronavirus~\citep{ahmed2021numerical}.

Motivated by the American COVID-19 data, we consider adding an advection term into the PDE in (\ref{wiklepde}) to capture the spatio-temporal trend of the spread. The advection-diffusion-reaction PDE is: 

\begin{equation}\label{eq12}
    \frac{\partial}{\partial t}u(t;s)=-\boldsymbol{\nu}(t;s) \nabla u(t;s)+\nabla \cdot \delta(s) \nabla u(t;s) +\zeta(t) u(t;s),
\end{equation}

\noindent where $\boldsymbol{\nu} (t;s)=(\nu_1(t;s),\nu_2(t;s))$. Here $\nu_1(t;s)$ and $\nu_2(t;s)$ measure the tendency of the virus to move in the positive direction of the x- and y-axis. The drift velocity and growth rate can be time-independent with $\boldsymbol{\nu} (t;s)\equiv(\nu_1(s),\nu_2(s))$ and $\zeta(t)\equiv\zeta$ for all $t$.
 
Denote $u_t(x,y)=u(t;s)|_{s=(x,y)}$, we approximate $u(t;s)$ in (\ref{eq12}) using the following equation:
\footnotesize
\begin{equation}\label{makeH}
\begin{aligned}
\tilde{u}_t(x,y) &= u_{t-\Delta t}(x,y)[1-2\delta(x,y)(\frac{\Delta t}{\Delta^2x}+\frac{\Delta t}{\Delta^2y})+\zeta_{t- \Delta t} \Delta t] \\
&+ u_{t-\Delta t}(x-\Delta x,y)[\frac{\Delta t}{4\Delta^2x}\{4\delta(x,y)-\delta(x+\Delta x,y)+\delta(x-\Delta x,y)+2\nu_{1,t-\Delta t}(x,y)\Delta x\}]\\
&+ u_{t-\Delta t}(x+\Delta x,y)[\frac{\Delta t}{4\Delta^2x}\{4\delta(x,y)+\delta(x+\Delta x,y)-\delta(x-\Delta x,y)-2\nu_{1,t-\Delta t}(x,y)\Delta x\}]\\
&+ u_{t-\Delta t}(x,y+\Delta y)[\frac{\Delta t}{4\Delta^2y}\{4\delta(x,y)+\delta(x,y+\Delta y)-\delta(x,y-\Delta y)-2\nu_{2,t-\Delta t}(x,y)\Delta y\}]\\
&+ u_{t-\Delta t}(x,y-\Delta y)[\frac{\Delta t}{4\Delta^2y}\{4\delta(x,y)-\delta(x,y+\Delta y)+\delta(x,y-\Delta y)+2\nu_{2,t-\Delta t}(x,y)\Delta y\}],\\
\end{aligned}
\end{equation}
\normalsize

\noindent where $s=(x,y)$ represents a rectangular region $[x-\frac{1}{2}\Delta x,x+\frac{1}{2}\Delta x]\times[y-\frac{1}{2}\Delta y,y+\frac{1}{2}\Delta y]$. It is an approximation of the solution of PDE by first-order forward differences in time and centred differences in space. 

We express $u(t;s)$ in vector form based on (\ref{makeH}):

\begin{equation}\label{eq2.6}
    u(t;s)=\boldsymbol{H}(\boldsymbol\delta,\zeta(t-\Delta t),\boldsymbol\nu(t-\Delta t))\boldsymbol u(t-\Delta t)+\epsilon(t;s),
\end{equation}

\noindent where $\boldsymbol{H}(\boldsymbol{\delta},\zeta(t-\Delta t),\boldsymbol{\nu}(t-\Delta t))$ is a $n$-dim vector, $\boldsymbol u(t-\Delta t)$ represents the vector with the spatio-temporal process at all the locations at the previous time point, and $\epsilon(t;s)$ is the error term and we assume $\epsilon (t;s) \sim N(0,\sigma^2)$ iid. Alternatively, we can consider errors with autoregressive structure:

\begin{equation}\label{autoepsilon}
\epsilon(t;s)= \phi  \epsilon(t-\Delta t;s)+\eta(t;s),
\end{equation}

\noindent and assume $\phi \sim U(-1,1)$, $\eta(t;s) \sim N(0,\sigma^2)$. 

In the COVID-19 application, we let $s \in \mathcal{S}=\{1,2,...,26\}$ as shown in Figure \ref{3.1} A. We consider evenly-spaced time points $t \in \mathcal{T}=\{1,2,...,23\}$, where $t=1$ and $t=23$ respectively represents April 2020 and Feb 2022. We set $\Delta t=1$ to be one month. Given that the spread over a long period is of interest, it is sufficient to use the monthly data. When more detailed temporal features are desired, smaller time units can be used at the cost of increased computation times.

\subsubsection{Prior Distribution of Parameters} 

Assume for any $t$ and $s$, the advection, diffusion and growth terms follow normal distribution:

\begin{equation}
\begin{aligned}
\nu_1(t;s) &\sim N(\tilde{\nu}_1,\tilde{\sigma}^2_{\nu_1}), & \delta(s)&\sim N(\tilde{\delta},\tilde{\sigma}^2_{\delta}),\\
\nu_2(t;s) &\sim N(\tilde{\nu}_2,\tilde{\sigma}^2_{\nu_2}), & \zeta(t)&\sim N(\tilde{\zeta},\tilde{\sigma}^2_{\zeta}).\\
\end{aligned}
\end{equation}

In addition, we let $\delta(s) \geq 0$, and $\sigma^2 \sim IG(\tilde{q},\tilde{r})$. The parameters with the “tilde” are hyperparameters that are previous specified. The choices for the hyperparameters can be based on the prior knowledge from the transmission of the infectious disease in other regions. In cases when this information is limited, we choose vague prior distributions.

\subsection{Estimation Procedure}\label{chap:2.3}

The following posterior distribution summarises the Bayesian formulation of the proposed hierarchical model:
\begin{equation}
\begin{aligned}
[ \boldsymbol{\nu}_{1,0},...,\boldsymbol{\nu}_{1,T-1},\boldsymbol{\nu}_{2,0}&,...,\boldsymbol{\nu}_{2,T-1},\boldsymbol{\delta},\zeta_0,...,\zeta_{T-1},\sigma^2, \phi, \boldsymbol u_1,...,\boldsymbol u_T \mid \boldsymbol n_1,...,\boldsymbol n_T ]\\
\propto\{\prod_{t=1}^T[\boldsymbol n_t \mid \boldsymbol u_t] \}
\times \{\prod_{t=1}^T[& \boldsymbol u_t \mid \boldsymbol \nu_{1,t-1},\boldsymbol \nu_{2,t-1},\boldsymbol \delta, \zeta_{t-1},\boldsymbol u_{t-1},\sigma^2,\phi]\}\\
 \times \  [ \boldsymbol{\nu}_{1,0}]...[\boldsymbol{\nu}_{1,T-1}][\boldsymbol{\nu}_{2,0}]&...[\boldsymbol{\nu}_{2,T-1}][\boldsymbol{\delta}],[\zeta_0]...[\zeta_{T-1}][\sigma^2][\phi].
\end{aligned}
\end{equation}

Since the posterior distribution is not analytically tractable, Markov chain Monte Carlo (MCMC) methods is applied to obtain samples from the posterior distribution of the parameters. The procedure was carried out using RStan~\citep{RStan} in R~\citep{Rsoftware}.

After the prior and the sampling distribution are specified, the Gibbs sampler cycles through the full conditional distribution for the parameters. After the Markov chains converge, each parameter's point estimate is obtained using its posterior mean.

\section{Numerical Studies}\label{chptnume}

\subsection{Simulation Study}

We conduct a simulation study to assess the proposed approaches numerically. In simulation settings, we set $T=24$ and $S=5\times 5=25$ grids. The true values of parameters are provided in Table \ref{hs1}, and the hyperparameters used in both scenarios are provided in Table \ref{prior}. In both the scenarios, the counts drift to the bottom-right corner for $t=1,...,12$ and drift in the opposite direction for $t=13,...,24$. Also, the growth rate is high initially and then drops to about 0.

\subsubsection{Scenario 1: Time-varying Growth Rate, Space-varying Diffusion Rate, and Time- and Space-varying Advection Speed}

We generated data based on the proposed model 3 and conducted analyses under the first four models as listed in Table \ref{table2}. The data generating process is as follows:

\begin{algorithm}[H]
\For{$t\gets1$ \KwTo $24$ }{
    Calculate vector $\boldsymbol{H}(\boldsymbol\delta,\zeta(t-1),\boldsymbol\nu(t-1))$ based on (\ref{makeH})\;
    \For{$s\gets1$ \KwTo $25$ }{
    Generate the errors: $\epsilon (t;s) \sim N(0,\sigma^2)$ iid\;
    Calculate the spatio-temporal process $u(t;s)$ following (\ref{eq2.6})\;
    Calculate the Poisson intensity process: $\lambda(t;s)=exp(u(t;s)$)\;
    Generate counts: $n(t;s) \sim Poisson(\lambda(t;s))$ \;
    }
    }
 \caption{Data Generating Process in Scenario 1}
\end{algorithm}

\vspace{5mm} 

We conducted the analyses of the generated data utilizing the four aforementioned models. By the MCMC method described in Section \ref{chap:2.3}, samples were obtained from the posterior distribution for each parameter. 50,000 MCMC samples were drawn, with the first 20,000 as the burn-in. The values of R-hats, a diagnostic statistics comparing the variation between chains to the variation within the chains, are all close to 1. It verifies the adequacy of the burn-in period.

The true values and the posterior means of the parameters and $\boldsymbol u(t;s)$ are shown in Figure 2 and 3 in the Supplementary Materials. The mean squared errors (MSEs) associated with each model are also shown in Figure 3 in the Supplementary Materials. The proposed model 3 outperforms the other three models with the closest estimates of the parameters and the lowest MSE values.

\subsubsection{Scenario 2: Process with Autoregressive Errors}\label{chpt:4.2}

In this scenario, we generate data from the proposed model 3 with autoregressive errors to examine the robustness of the estimation procedure. We kept the same true values for the parameters as in the first scenario.

The data generating process was the same as the steps in scenario 1 with exception in the generation of $\epsilon(t;s)$. We assume 
\begin{equation}
\epsilon(t;s)= \phi  \epsilon(t-1;s)+\eta(t;s),
\end{equation}
where $\eta(t;s) \sim N(0,0.1)$, and $\phi=0.1$. These values were selected since they are close to the COVID-19 analysis results in the following section.

The true values and the estimation of parameters for all the models are shown in Figure \ref{sim2par}. Figure \ref{sim2u} illustrates the true values and the estimated values of $\boldsymbol u_t$ with all the models. The proposed model 3 again produces a reasonable estimation of parameters and $\boldsymbol u_t$, indicating its robustness to model misspecification.

We consider four different prior variances for advection, diffusion and reaction rates for proposed model 3. Figure 4 in Supplementary Materials shows the estimated parameters for different prior distributions. The estimates are close to the true values when the variance is less than or equal to 0.01. It indicates certain robustness of the analysis under the proposed model 3. However, estimates can differ from the true values when the prior variances are large, which motivates the proposed model 5, specifying correlations between the random errors.

\subsection{Data Analysis: COVID-19 Infection}\label{chpt3dataanalysis}

We apply the proposed approach to investigate the spread of the COVID-19 infection in the USA from Mar 2020 to Mar 2022 using the data file downloaded from \emph{The New York Times}.

\subsubsection{Monthly Confirmed Cases Aggregated by Grids}

We divide the mainland USA into 26 equal-area grid cells, with each covering $174,807$ miles square, as shown in Figure \ref{3.1}. The size of the cells depends on the selection of $\Delta x$ and $\Delta y$ in (\ref{makeH}). Higher resolution grids can be used if more detailed spatial features are desired, at the expense of longer computation times. Figure \ref{3.3} A presents log-transformed aggregated counts by the grid using the geographic coordinates for each county from US Census Bureau~\citep{us-census-bureau-2021}. This plot again shows that the spatio-temporal process involves great variability over space and time. Figure \ref{3.1} B illustrates how monthly cases vary in each grid cells, demonstrating that the counts have an uncertain tendency over time. Since population sizes vary across cells, the counts alone cannot accurately reflect the level of disease. Population-adjusted case maps and line plots are shown in Figure \ref{3.3} B and Figure \ref{3.1} C, respectively. This allows us to compare the level of disease in different populations.

\subsubsection{Bayesian Analysis: Spread of COVID-19 with Hierarchical Bayesian Model}\label{chap:3.2}

We consider six different models specified in Table \ref{table2} to investigate the spread of COVID-19 in the USA. 
In proposed model 5, we consider both autoregressive structure error and population-adjusted latent process. We account for the spatial variations in the control measures and other factors using space-varying diffusion rate and drift velocity. The growth rate is assumed space-independent based on the findings from Figure \ref{3.3} B. This is because the potential major factors, such as the availability of vaccines and the emergence of variants, are similar across the country. In addition, we consider the drift velocity and growth rate to be time-varying since there is great variability of case counts over time according to the COVID-19 data, as shown in Figure \ref{3.1} C. Given that most of the temporal spread is modelled by these two parameters, the additional time-varying feature of the diffusion rate is unnecessary.

Other models are nested in model 5. The monthly confirmed cases from Mar 2020 to Feb 2022 are used as input data. After obtaining the parameter estimates, we make predictions for case numbers in Mar 2022 and compare them to the real counts from \emph{The New York Times}.

Table \ref{table3} gives the hyperparameter values used in all of our analyses. Given the limited information on the parameters, we choose vague priors by forcing the variances to be relatively large. It is verified that the analyses are not sensitive to the selections of the hyperparameters by testing with three different sets of prior values listed in Table \ref{table3}.

We run three Markov chains for each parameter using RStan. $300,000$ samples are drawn from the posterior distribution from each chain, with the first $150,000$ considered burn-in. The R-hats are close to 1, indicates that the chains converged. 

Figure \ref{5} illustrates the estimate of the parameters with their posterior mean under each model. The growth rate in proposed model 2-5 fluctuate over time and follows a similar pattern as presented in Figure 1 in the Supplementary Materials. The spatial map of the posterior mean for the diffusion rate, $\delta$, suggests significant heterogeneity in the diffusion of the latent process. Figure \ref{5} also includes the estimated $\nu(t;s)$, the trend of spread, where the size of the arrow is proportional to the value of advection speed in each grid. The analysis results confirm that the advection velocity are time-varying. In proposed model 5, the 95\% credible interval for $\phi$ is $[0.231,0.238]$ indicates that it is significantly different from 0.

We make predictions for case numbers in Mar 2022 following the steps below. Here, $t=23$ and $t=24$ represents Feb 2022 and Mar 2022, respectively.

\begin{algorithm}[H]
\For{$t=23$ }{
    Obtain posterior mean for each of $\phi$, $\boldsymbol\delta$, $\boldsymbol{\epsilon}(t)$ $\zeta(t)$, $\boldsymbol\nu_1(t)$, $\boldsymbol\nu_2(t)$ and $\boldsymbol u(t)$\;
    Calculate matrix $\boldsymbol H(\boldsymbol\delta,\zeta(t),\boldsymbol \nu_1(t),\boldsymbol \nu_2(t))$ based on (\ref{makeH});
    }
\For{$t=24$ }{
    Estimate $\boldsymbol u(t)=\boldsymbol{H}\boldsymbol{u}(t-1)+\phi  \boldsymbol \epsilon(t-1)+log(\boldsymbol p)$, where $\boldsymbol p$ is a vector of population;\
    }
 \caption{Prediction Procedure for Case Numbers in Mar 2022}
\end{algorithm}

\vspace{5mm} 

Figure \ref{7} illustrates the difference between the log-transformed observed cases and the prediction of the log-transformed cases in Mar 2022. The proposed models with time-varying $\zeta$ outperform other models since they have closer estimations to the true values of parameters.

\section{Discussion}\label{chptdiscussion}
This paper aims to develop a tool to explore the spatio-temporal spread of events. We attempt to understand the spread features of events and make predictions about the future. Our modeling is partly motivated by the American coronavirus infection data. We introduce an advection-diffusion-reaction PDE with space- and time-varying parameters to a hierarchical Bayesian framework. Along with FDM and MCMC methods, we implement the procedure and estimate the parameters in the models. The proposed approach can be applied to analyze other types of spatio-temporal data, for example, the number of deaths or hospitalizations due to COVID-19 infection. 

We consider an additional advection term and the time-varying feature of parameters based on the model proposed by Wikle. These modifications play important roles on understanding the spread of events as shown in the prediction of American COVID-19 confirmed cases in Mar 2022. The parameter estimates are far away from the true values if we consider original model.

There are a few worthwhile future investigations along the line of dependence of the key quantities on time, location and other exposures. This paper considers the growth rate and advection velocity as piecewise constant. As we mentioned in Section \ref{chap:1.2}, the trend of case counts is largely affected by many factors, such as the control measures or the emergence of the variants of coronavirus. We could further modify the model by considering the key quantities as the functions of time, location, and other potential covariates.

The error term is assumed to be independent across space in the process modeling. However, this assumption might be too strong. The counts are likely correlated with each other spatially. Future investigations are still needed for understanding the impact of possible spatial dependence.

Other future investigations includes: evaluating the process given parameters rather than generating the process from the joint distribution of parameters and process; choosing smaller or dynamic partition for $\Delta x$, $\Delta y$ and $\Delta t$ based on the data; considering alternative boundary conditions~\citep{wikle2003hierarchical}.

\newpage
\begin{appendix}

\section{Tables}
\begin{table}[h]
  \begin{center}
      \caption{The True Values of Parameters Used in Both Simulation Scenarios \label{hs1}}
    \begin{tabular}{cc|cc} 

      \textbf{parameter} & \textbf{true value} & \textbf{parameter} & \textbf{true value}\\ \midrule
      
      $\delta(s)$ for $s=1,...,25$ & 0.1 & $\sigma^2 $  & 0.1\\
      $\zeta(t)$ for $t=0,...,5$ & 0.15 & $\zeta(t)$ for $t=6,...,23$ & 0.01\\
      $\nu_1(t)$ for $t=0,...,11$ & 0.1 & $\nu_1(t)$ for $t=12,...,23$ & -0.1\\
      $\nu_2(t)$ for $t=0,...,11$ & 0.1 & $\nu_2(t)$ for $t=12,...,23$ & -0.1\\

    \end{tabular}
  \end{center}
\end{table}
\begin{table}[h]
  \begin{center}
    \caption{Hyperparameters Used in Both Simulation Scenarios\label{prior}}
    \begin{tabular}{cc|cc} 

      \textbf{parameter} & \textbf{true value} & \textbf{parameter} & \textbf{true value}\\ \midrule

      $\tilde{\nu}_{1}$ & 0 & $\tilde{\nu}_{2}$ & 0\\
      $\tilde{\sigma}^2_{\nu_1}$ & 0.1 & $\tilde{\sigma}^2_{\nu_2}$ & 0.1\\
      $\tilde{\delta}$ & 0 & $\tilde{\zeta}$ & 0\\
      $\tilde{\sigma}^2_\delta$ & 0.1 & $\tilde{\sigma}^2_\zeta$ & 0.1\\
      $\tilde{q}$ &0.001 & $\tilde{r}$ & 0.001\\

    \end{tabular}
  \end{center}
\end{table}  
\begin{table}[h]
  \begin{center}
      \caption{Description of Models \label{table2}}

    \begin{tabular}{c|cccccccc} 

      \textbf{Models}  & \textbf{TV  $\zeta$} & \textbf{SV  $\zeta$}& \textbf{TV  $\delta$} &\textbf{SV  $\delta$} & \textbf{TV  $\nu$}& \textbf{SV  $\nu$}& \textbf{PA}& \textbf{AR  $\epsilon$}\\ \midrule
      Wikle & \xmark & \xmark & \xmark &\cmark & / &/&\xmark &\xmark \\

      Proposed Model 1 & \xmark & \xmark & \xmark &\cmark & \xmark &\cmark&\xmark &\xmark \\

     Proposed Model 2 &  \cmark & \xmark & \xmark &\cmark & \xmark &\cmark&\xmark &\xmark \\

     Proposed Model 3 & \cmark & \xmark & \xmark &\cmark & \cmark &\cmark&\xmark &\xmark  \\

      Proposed Model 4 & \cmark & \xmark & \xmark &\cmark & \cmark &\cmark&\cmark &\xmark  \\

            Proposed Model 5 & \cmark & \xmark & \xmark &\cmark & \cmark &\cmark&\cmark &\cmark  \\\midrule

      \multicolumn{9}{l}{\footnotesize Notes: TV, time-varying; SV, space-varying; PA, population-adjusted; AR, autoregressive.} \\
    \end{tabular}
  \end{center}
\end{table}
\begin{table}[h]
  \begin{center}
    \caption{Values of the Hyperparameters\label{table3}}

    \begin{tabular}{c|ccc} 

      \textbf{parameter} & \textbf{Set 1} & \textbf{Set 2} & \textbf{Set 3}\\ \midrule

      $\tilde{\nu}_{1}$ & 0 & 0 & 0.1\\
      $\tilde{\nu}_{2}$ & 0 & 0 & 0.1\\
      $\tilde{\sigma}^2_{\nu_1}$ & 10& 100 &10\\
      $\tilde{\sigma}^2_{\nu_2}$ & 10& 100 &10\\
      $\tilde{\delta}$ & 0 & 0 & 0.1\\
      $\tilde{\zeta}$ & 0 & 0 & 0.1\\\
      $\tilde{\sigma}^2_\delta$ & 10 & 100 &10\\
      $\tilde{\sigma}^2_\zeta$ & 10&100 &10\\
      $\tilde{q}$ & 0.001 & 0.001 & 0.001\\
      $\tilde{r}$ & 0.001 & 0.001 & 0.001\\

    \end{tabular}
  \end{center}
\end{table}

\newpage
\section{Figures}

\begin{figure}[h]
\centering
        \includegraphics[width=14cm]{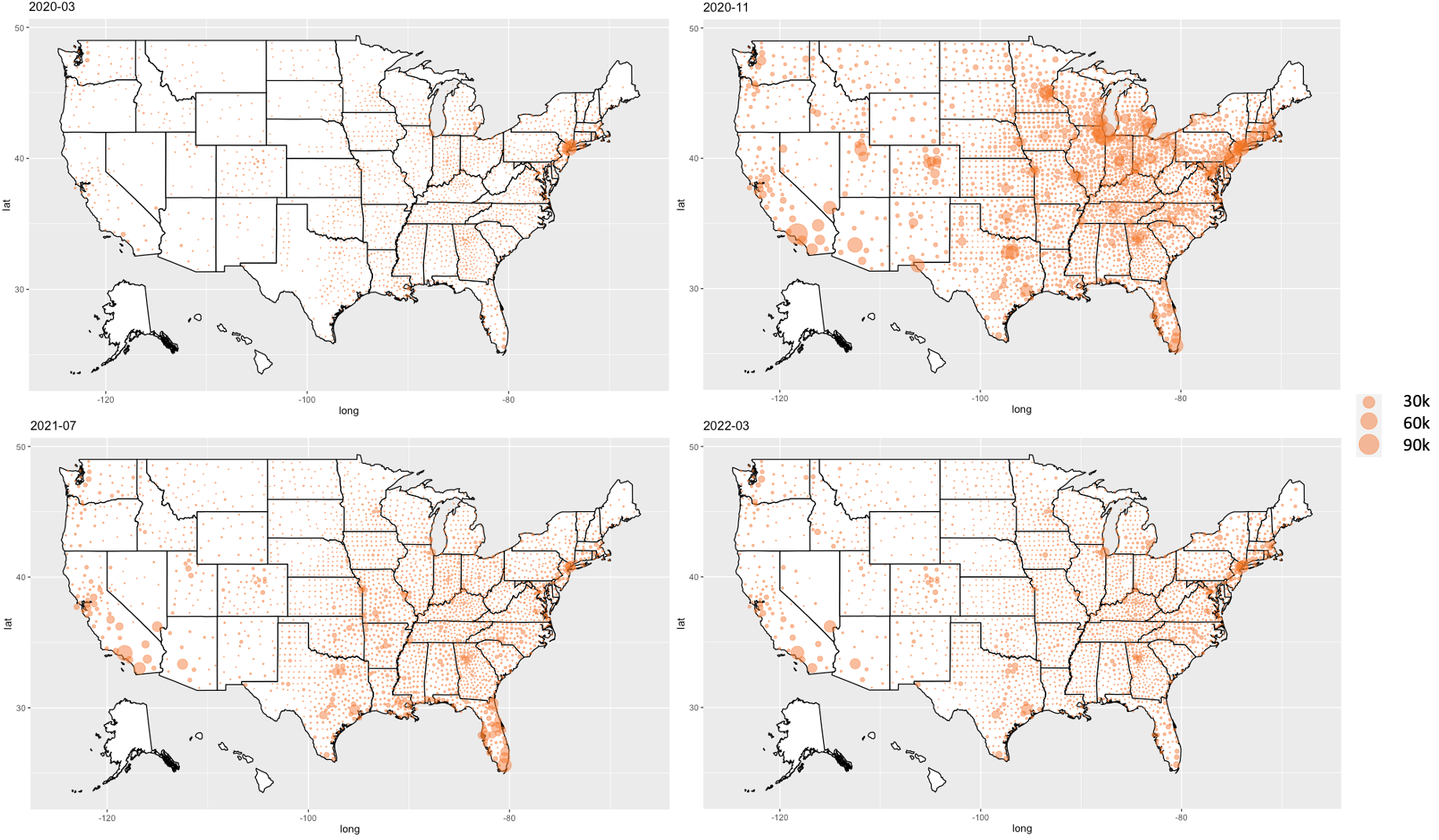}
  \caption{\label{1.1}The Confirmed COVID-19 Cases in the USA for Selected Months} 
\end{figure}

\begin{figure}[h]
\centering
        \includegraphics[width=14cm]{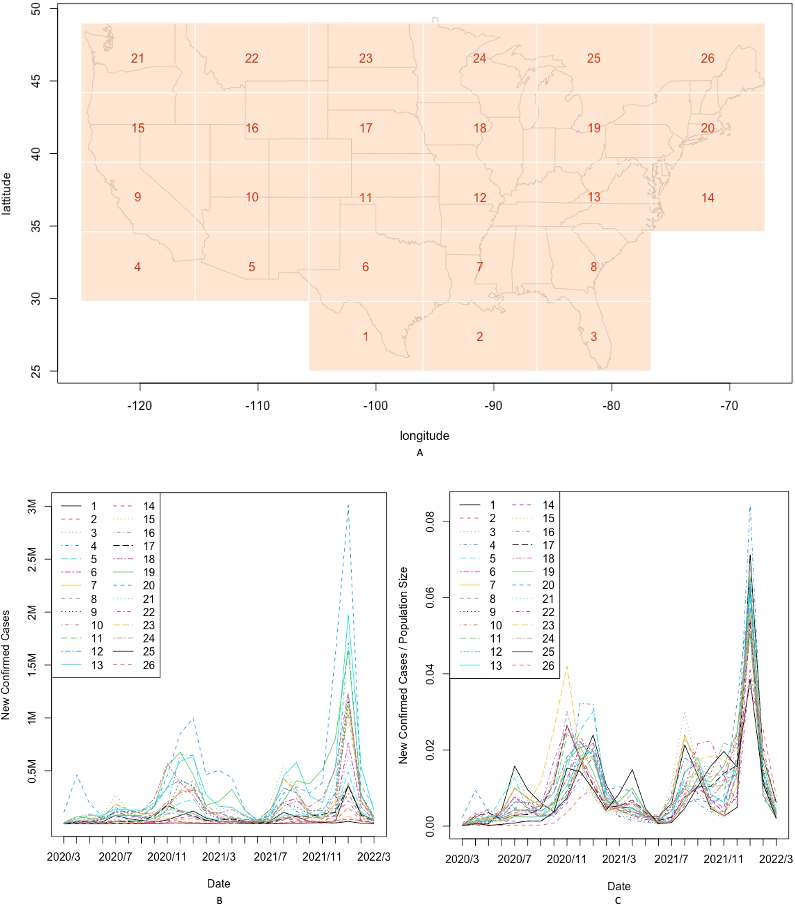}
  \caption{\label{3.1}A. Grids with the Partition in the Example: Location $s=1,...,S=26$; B. Monthly Confirmed New Cases; and C. Monthly Confirmed New Cases divided by the Population Sizes across Different Locations} 
  
\end{figure}

\begin{figure}[h]
\centering
        \includegraphics[width=12cm]{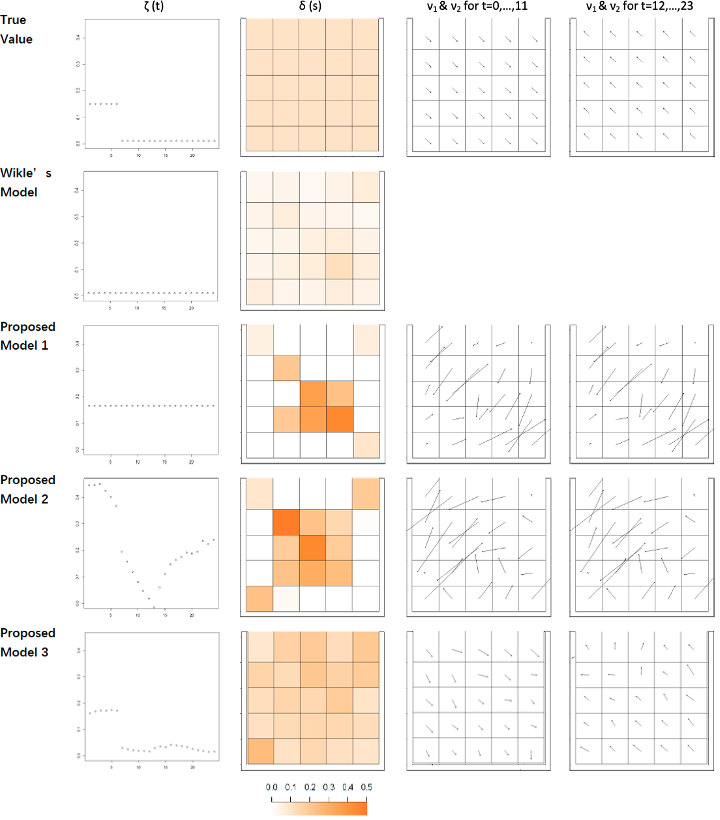}
  \caption{The True Values and the Posterior Means of Parameters with Different Models in Simulation Scenario 2\label{sim2par}}
\end{figure}

\begin{figure}[h]
\centering
        \includegraphics[width=13cm]{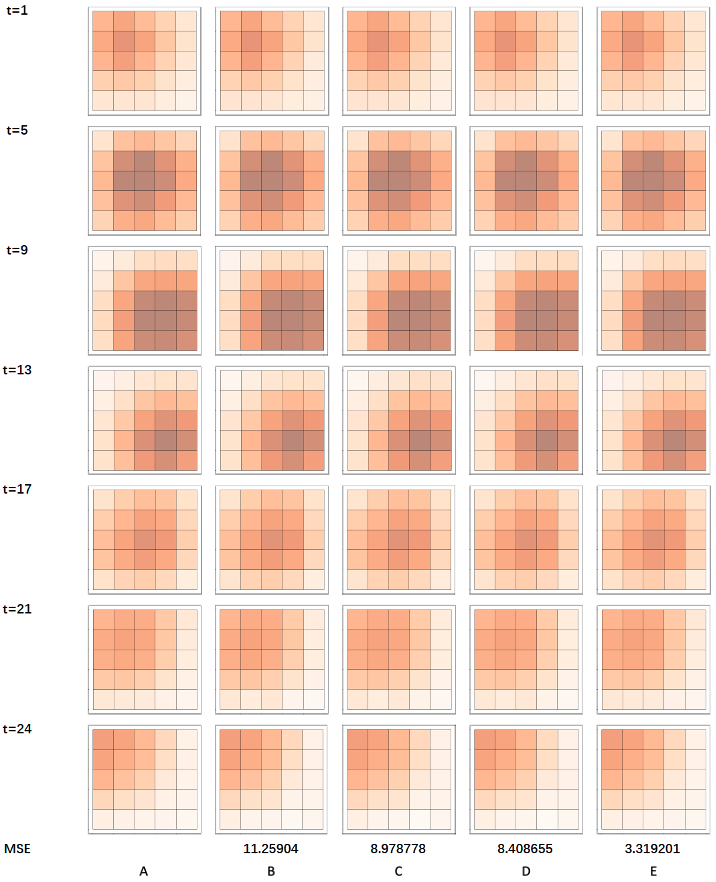}
  \caption{A. The True Values of $\boldsymbol u_t$; the Estimated Values of $\boldsymbol u_t$ with B. Wikle's Model, C. Proposed Model 1, D. Proposed Model 2 and E. Proposed Model 3 in Simulation Scenario 2 \label{sim2u}}
\end{figure}

\begin{figure}[h]
\centering
        \includegraphics[width=13cm]{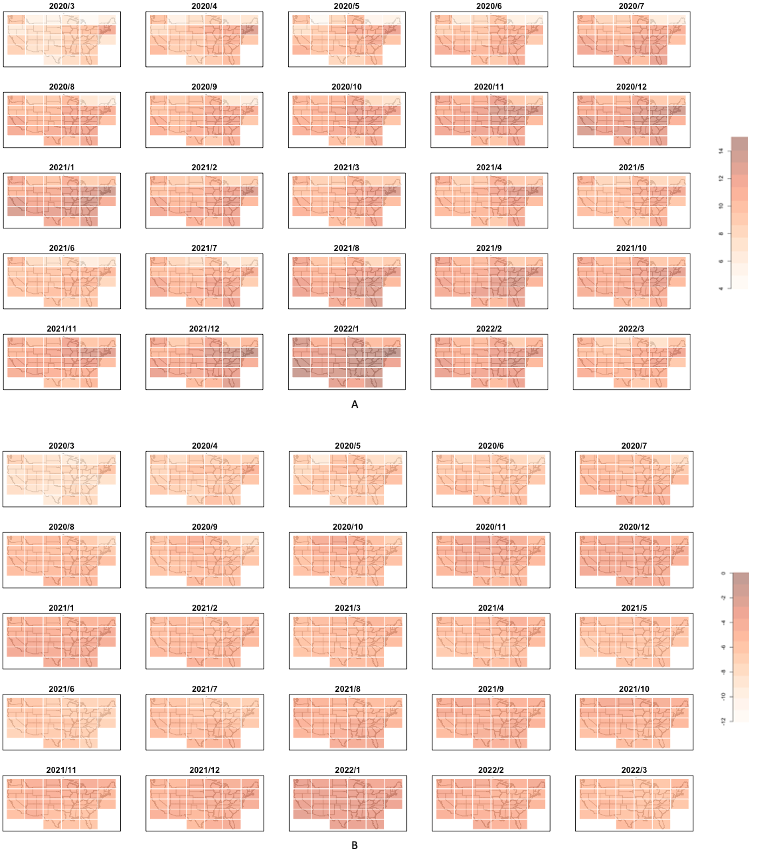}
  \caption{\label{3.3}A. Log-transformed Number of Confirmed Cases Aggregated by Grids; B. Population-Adjusted Log-transformed Number of Confirmed Cases Aggregated by Grids} 
\end{figure}

\begin{figure}[h]
\centering
        \includegraphics[width=13cm]{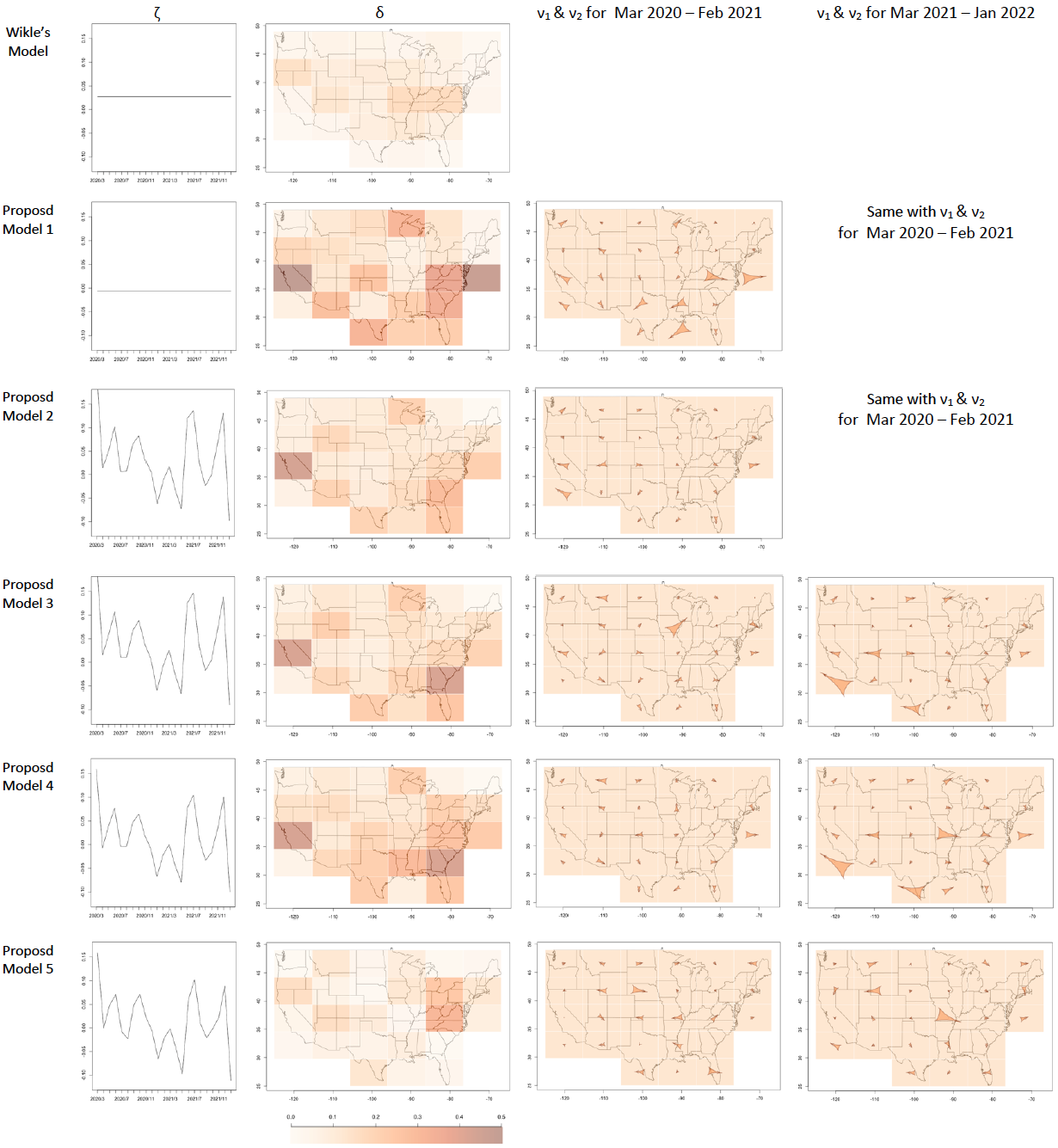}
  \caption{Posterior Means of Parameters in Each Model\label{5} } 
  
\end{figure}

\begin{figure}[h]
\centering
        \includegraphics[width=12cm]{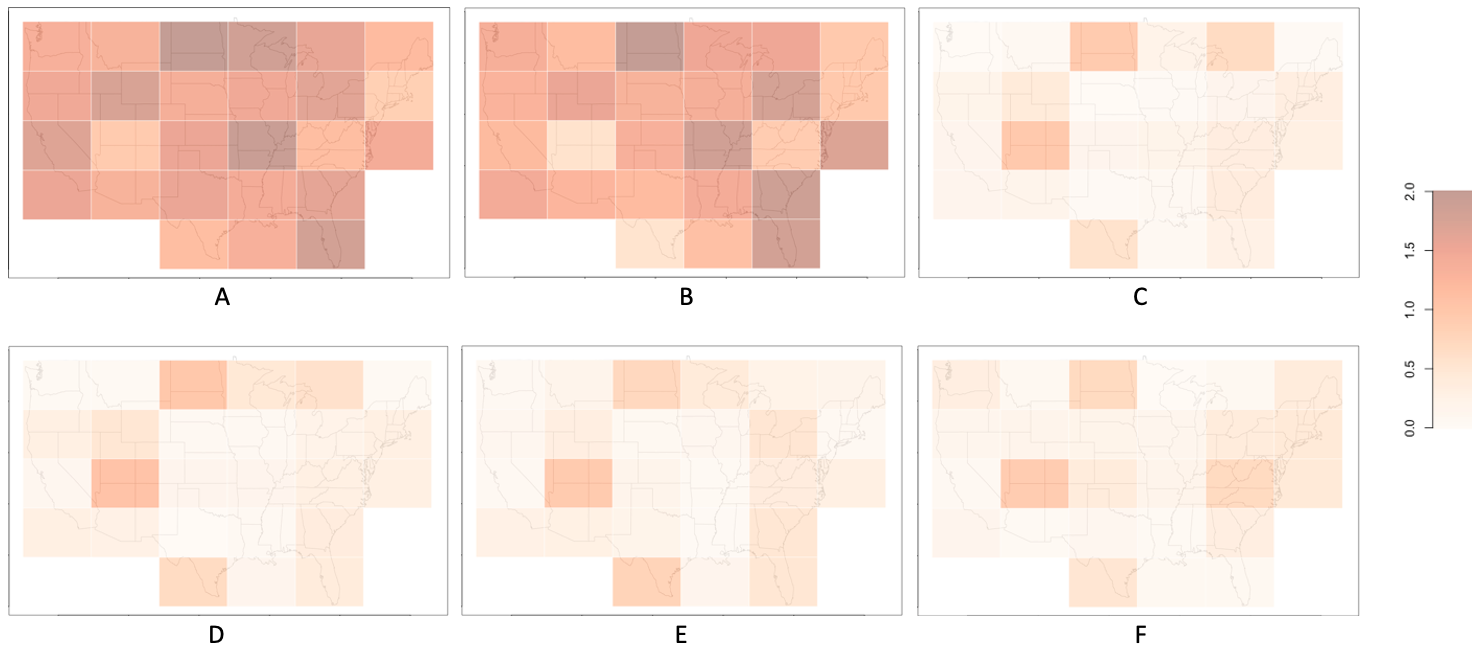}
  \caption{Difference between Log-transformed Observed Counts in Mar 2022 with Prediction of Log-transformed Counts under A. Wikle's Model, B. Proposed Model 1, C. Proposed Model 2, D. Proposed Model 3, E. Proposed Model 4 and F. Proposed Model 5\label{7} } 
  
\end{figure}

\end{appendix}

\clearpage 

    \bibliographystyle{plainnat}
    \bibliography{references}

\newpage

\section*{Supplementary}
\subsection*{1  Additional Descriptive Analysis}
\begin{figure}[h]
\centering
        \includegraphics[width=14cm]{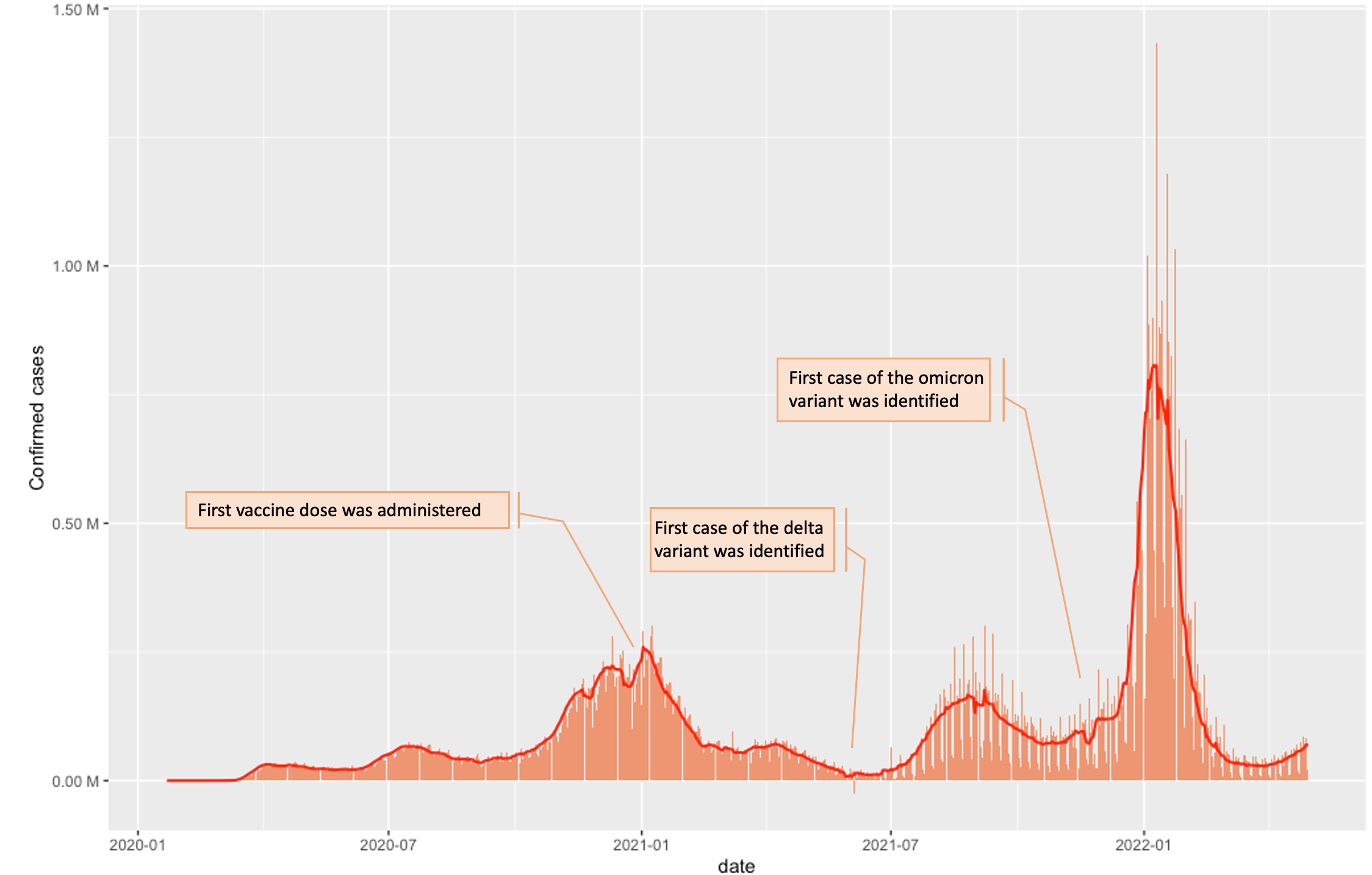}
  \caption{\label{1.2}Daily Confirmed New Cases and 7-day Moving Average in the USA} 
\end{figure}

Figure \ref{1.2} presents the daily cases with a 7-day moving average in the USA. There were three large spikes in the confirmed case numbers over the past two years. The first one occurred during the holiday season in 2020. Gatherings and celebrations accelerated the spread of the disease. In June 2021, the number of cases fell to the lowest point of the year. This decline was likely associated with the availability of vaccines. The second and third sharp increases were consistent with the emergence of delta and omicron variants. The data suggest that these variants are more transmissible than the original virus~\citep{CDC}. Therefore, we can further modify the model by incorporating the key quantities as functions of potential covariates.

\subsection*{2  Additional Simulation results}
\begin{figure}[h]
\centering
        \includegraphics[width=13cm]{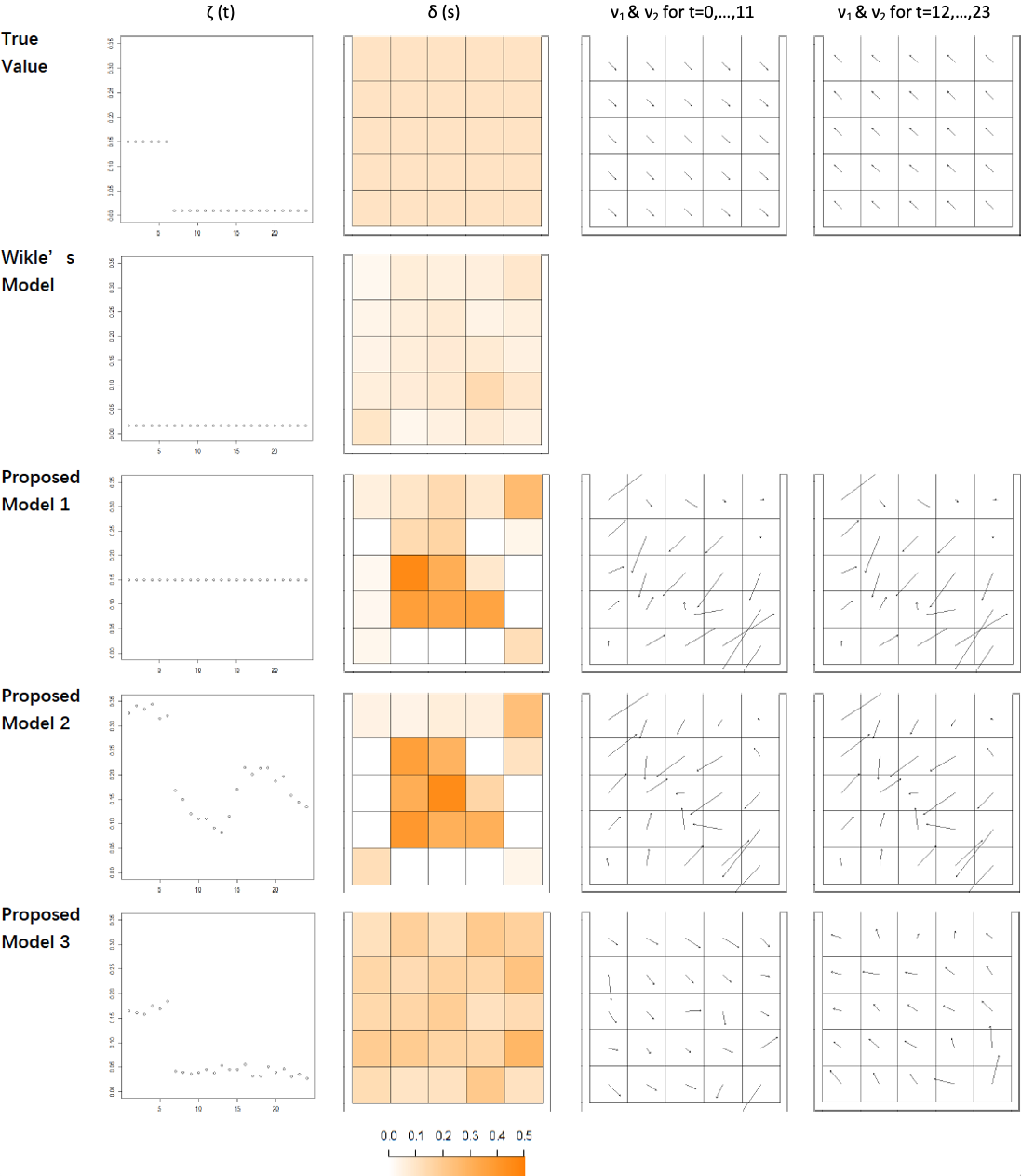}
  \caption{The True Values and the Posterior Means of Parameters from Different Models in Simulation Scenario 1 \label{sim1par}}
\end{figure}

\begin{figure}[h]
\centering
        \includegraphics[width=12cm]{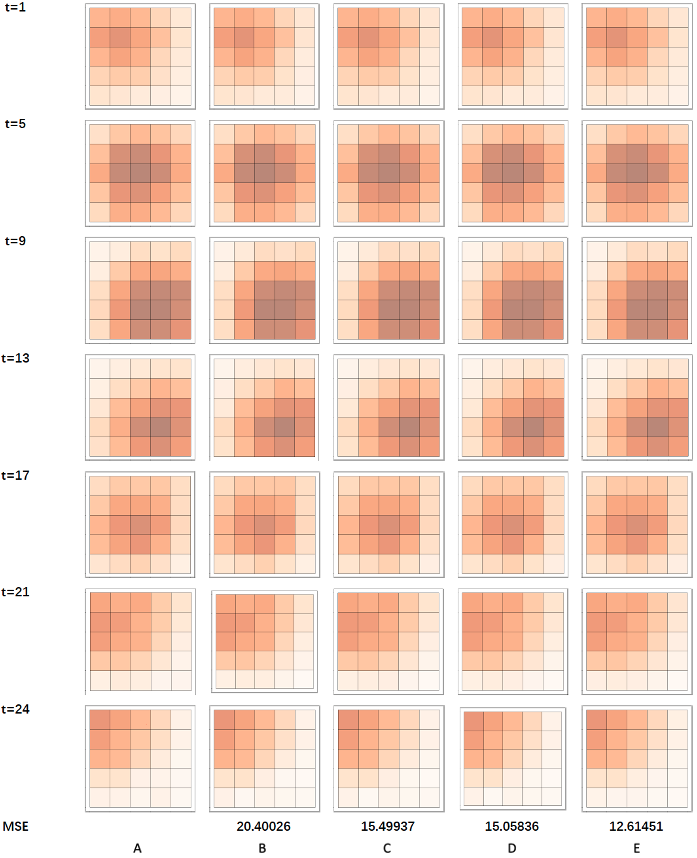}
  \caption{A. The True Values of $\boldsymbol u_t$; the Estimated Values of $\boldsymbol u_t$ with B. Wikle's Model, C. Proposed Model 1, D. Proposed Model 2 and E. Proposed Model 3 in Simulation Scenario 1 \label{sim1u}}
\end{figure}

Figures \ref{sim1par} and \ref{sim1u} show the true values of the parameters and $\boldsymbol u(t;s)$ in simulation scenario 1, as well as their posterior means under Wikle's and proposed models. The proposed model 3 outperformed the others, with the closest estimation of parameters.

\begin{figure}[h]
\centering
        \includegraphics[width=13cm]{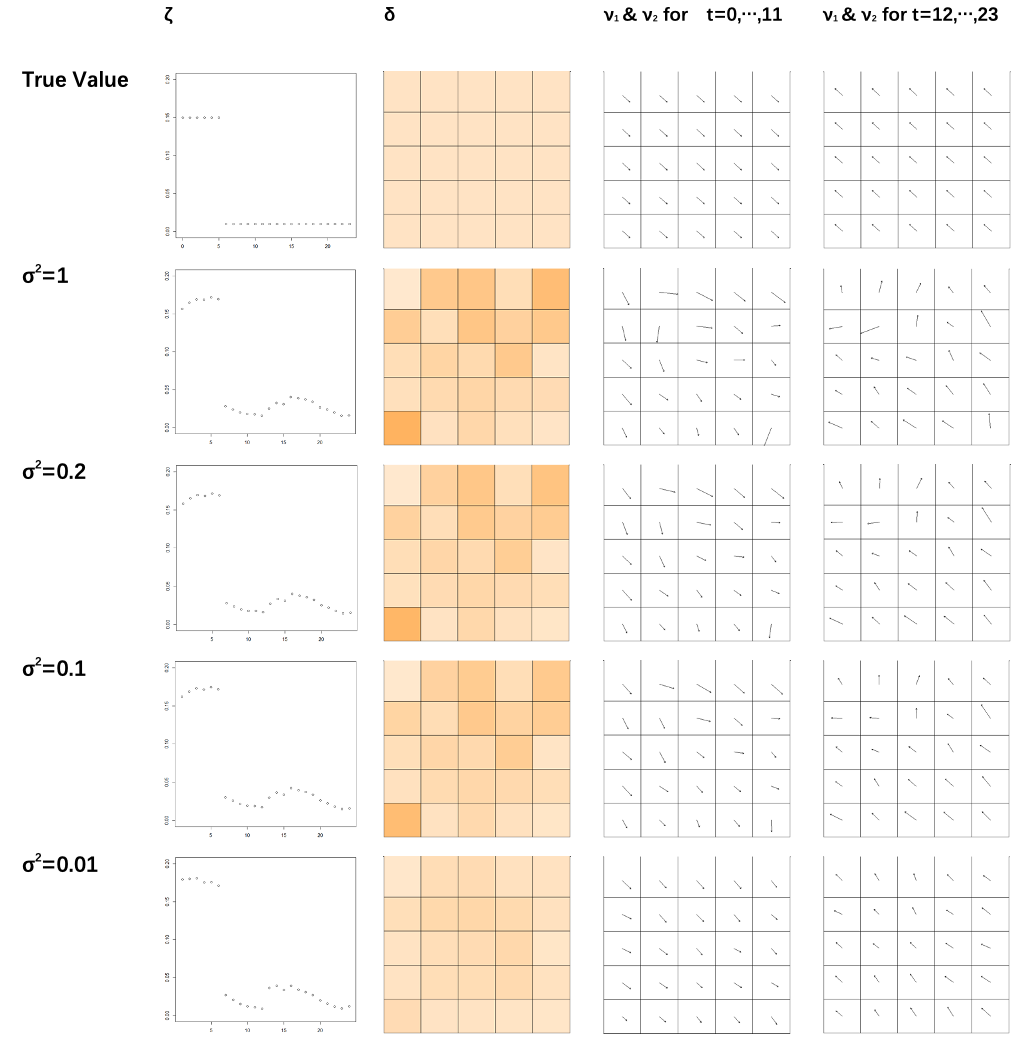}
  \caption{The True Values and the Posterior Means of Parameters with different prior distributions in Simulation Scenario 2 \label{diffprior}}
\end{figure}

\newpage
For the proposed model 3, we examine four different prior variances for advection, diffusion, and reaction rates. The estimated parameter values are presented in Figure \ref{diffprior}. Based on the plot, it is evident that for variances less than or equal to 0.1, the parameters' estimates are close to their true values. The results indicate that the proposed model 3 is relatively robust. To resolve the problem of estimates deviating from the true values when the prior variances are large, we specify the correlations between random errors in the proposed model 5.

\end{document}